\documentclass[11pt]{article}

\usepackage{amsmath}
\usepackage{amsfonts}
\usepackage{graphicx}
\usepackage{bm}
\usepackage{algorithm}
\usepackage{algorithmic}
\usepackage{natbib}
\usepackage{float}
\usepackage{fullpage}

\newcommand{\bQ}[0]{\bm{Q}}
\newcommand{\bx}[0]{\bm{x}}
\newcommand{\by}[0]{\bm{y}}
\newcommand{\bA}[0]{\bm{A}}
\newcommand{\beps}[0]{\bm{\epsilon}}
\newcommand{\bmu}[0]{\bm{\mu}}
\newcommand{\mcN}[0]{\mathcal{N}}
\newcommand{\etab}[0]{\bm{\eta}}
\newcommand{\thetab}[0]{\bm{\theta}}
\newcommand{\bL}[0]{\bm{L}}
\newcommand{\be}[0]{\bm{e}}
\newcommand{\bv}[0]{\bm{v}}
\newcommand{\bI}[0]{\bm{I}}

\newcommand{\mcK}[0]{\mathcal{K}}
\newcommand{\mcO}[0]{\mathcal{O}}
\newcommand{\mbR}[0]{\mathbb{R}}
\newcommand{\tr}[0]{\textnormal{tr}\,}
\newcommand{\diag}[0]{\textnormal{diag}}

\DeclareMathOperator*{\argmax}{arg \, max}

\title{Parameter Estimation in High Dimensional Gaussian Distributions}

\author{Erlend Aune (NTNU, Norway) and Daniel P. Simpson (NTNU, Norway)}

\begin{document}

\maketitle

\begin{abstract}
In order to compute the log-likelihood for high dimensional spatial Gaussian models, it is necessary to compute  the determinant of the large, sparse, symmetric positive definite precision matrix, Q.  Traditional methods for evaluating the log-likelihood for very large models may fail due to the  massive memory requirements.  We present a novel approach for evaluating such likelihoods when the matrix-vector product, Qv, is fast to compute. In this approach we utilise matrix functions, Krylov subspaces, and probing vectors to construct an iterative method for computing the log-likelihood. 
\end{abstract}

\section{Introduction}

In computational and, in particular, spatial statistics, increasing possibilities for observing large amounts of data leaves the statistician in want of computational techniques capable of extracting useful information from such data. Large data sets arise in many applications, such as modelling seismic data acquisition (\cite{bul_omr_linavo}); analysing  satellite data for ozone intensity, temperature and cloud formations(\cite{mcpeters1996nimbus}); or constructing  global climate models (\cite{lindgren2010explicit}).  Most models in spatial statistics are based around multivariate Gaussian distributions, which has probability density function
\begin{align*}
	p(\bx) = (2\pi)^{-k/2} \det(\bQ)^{1/2} \exp \left( -\frac{1}{2}(\bx-\bmu)^T \bQ(\bx-\bmu) \right)
\end{align*} 
where the precision matrix $\bQ$ is the inverse of the covariance matrix.  In this paper, we assume that the precision matrix is sparse, which essentially enforces a Markov property on the Gaussian random field.  These models have better computational properties than those based on the covariance, and there are modelling reasons to prefer using $\bQ$ directly (\cite{lindgren2010explicit}).  We note that \cite{rue_tje_fitgmrf} showed that  it is possible to approximate general Gaussian random fields by Gaussian Markov random fields (GMRFs), that is Gaussian random vectors with sparse precision matrices.

Throughout this paper, we will consider the common  Gauss-linear model, in which our data is a noisy observation of a linear transformation of a true random field, that is 
\begin{align} \label{Gauss_linear}
	\by=\bA(\thetab) \bx + \beps_1,
\end{align}
where the matrix $\bA(\thetab)$ models the observation of the true underlying field $\bx$, known as the `forward model', while $\beps \sim \mcN(\bm{0},\bQ^{-1}_1)$ is the observation noise.  In order to complete the model, we require a prior distribution on $\bx$, which we take to be  Gaussian with mean $\bmu$ and  precision matrix $\bQ_x(\etab)$, and  appropriate hyperpriors are given for the mean $\bmu$, the hyperparameters in the prior $\etab$ and the hyperparameters in the forward model $\thetab$.

Given  a set of data, we wish to infer $\etab,\thetab$ and $\bx$. The ways in which this is done can vary, but in the end, several determinant evaluations are needed. One way to do this is to alternately estimate $\bx$ and $\etab,\thetab$ using the distributions $p(\bx|\by,\etab,\thetab)$ and $p(\etab,\thetab|\bx,\by)$, updating each consecutively. That is
\begin{enumerate}
	\item Find $\argmax_{\bx}p(\bx|\by,\etab,\thetab)$
	\item Find $\argmax_{\etab,\thetab}p(\etab,\thetab|\bx,\by)$
	\item Repeat until convergence.
\end{enumerate}
In a Gauss-linear model, the first step involves a linear solve, while the second is an optimisation over the space in which $\etab,\thetab$ reside. The distribution is
\begin{align}
	p(\etab,\thetab|\bx,\by) & \propto p(\by|\bx,\etab,\thetab)p(\etab,\thetab|\bx) \notag  \\
		& = p(\by|\bx,\thetab) p(\bx|\etab)p(\etab)p(\thetab) \label{eq_post_hyper}
\end{align}
The log of the last line in \eqref{eq_post_hyper} gives the objective function for the hyper-parameters in this case:
\begin{align}
	\Phi(\etab,\thetab) & =-\log p(\by|\bx,\thetab) p(\bx|\etab)p(\etab)p(\thetab) \label{eq_obj_hyper} 
\end{align}
We also have that $p(\by|\bx,\thetab)p(\bx|\etab)$ as a function of $\bx$ is $\mcN(\bmu_p,\bQ_p)$, where $\bQ_p=\bQ_x(\etab) + \bA(\thetab)^T \bQ_1 \bA(\thetab)$ and $\bmu_p = \bQ_p^{-1}(\bQ_x \bmu + \bA^T \bQ_1 \by)$. The separated expression, $p(\by|\bx,\thetab)p(\bx|\etab)$ in $\Phi$, is usually preferable, but this is problem dependent.

Expanding \eqref{eq_obj_hyper}, we get
\begin{align}
	\Phi(\etab,\thetab)= & C - \log(\frac{1}{2} \det(\bQ_1)) + \frac{1}{2} (\by - \bA(\thetab) \bx)^T \bQ_1 (\by - \bA(\thetab) \bx) -  \notag \\
		& \log(\frac{1}{2} \det(\bQ_x(\etab))) + (\bx-\bmu)^T \bQ_x (\bx-\bmu) - \log p(\etab) - \log p(\thetab). \label{eq_obj_exp}
\end{align}
In this expression, the only term which is difficult to evaluate is $\log \det ( \bQ_x(\etab))$, and it is needed for estimating the hyper-parameters, both for point estimates and for Gaussian- or Laplace-approximations of the hyper-parameters (see \cite{carlin2000bayes} for details on such approximations). It is this evaluation and its use in optimisation we address in this article.

\section{Determinant evaluations}

The most common way to compute the log-determinant of a  sparse precision or covariance matrix is to 1) reorder $\bQ$ to optimise for Cholesky factorisation, 2) perform a Cholesky factorisation of the reordered matrix $\bQ = \bL \bL^T$, 3) extract the diagonal entries of $\bm{l}=\textnormal{diag}(\bL)$ and 4) set the log-determinant as $\log \det \bQ = \sum_{j=1}^n 2 \log(l_j)$ (this comes from the identity $\det \bQ = \det \bL \det \bL^T = (\det \bL)^2$). The algorithm takes very few lines to program, given a good sparse matrix sorting routine, such as METIS (\cite{karypis1999fast}) and a fast sparse Cholesky factoring implementation, such as CHOLMOD (\cite{davis1999modifying},\cite{chen2008algorithm}). Problems occur, however, when there are massive amounts of fill-in in the Cholesky factorisation even after resorting the matrix in question. For a Gaussian Markov random field the dimensionality of the underlying parameter space affect the storage cost for computing the Cholesky factorisation. In $\mbR^1$ the cost is $\mcO(n)$, in $\mbR^2$, $\mcO(n^{3/2})$ in $\mbR^3$, $\mcO(n^2)$ (\cite{rue_gmrf}).

\subsection{Alternative approximations}

The starting point for an alternative, less memory intensive procedure for computing the log-determinant comes from the identity
\begin{align}
	\log \det \bQ = \tr \log \bQ \label{eq_logdet_ident}
\end{align}
In the symmetric positive definite case, this identity is proved noting that $\det \bQ=\prod_{i=1}^n \lambda_i$ where $\{\lambda_i\}$ are the eigenvalues of $\bQ$ and that $\log \bQ = \bm{V} \log(\bm{\Lambda}) \bm{V}^T$ with $\bm{\Lambda}=\diag(\bm{\lambda})$ and $\bm{V}$ contains the eigenvectors of $\bQ$. Furthermore, $\tr(\bm{V}\log(\bm{\Lambda})\bm{V}^T) = \tr(\bm{VV}^T \log\bm{\Lambda})=\tr \log \bm{\Lambda}$, which gives the identity.

How can this be useful in computation, is the next question. A trivial observation shows that
\begin{align*}
	\tr \log \bQ = \sum_{j=1}^n \be_j^T \log (\bQ) \be_j
\end{align*}
where $\be_j=(0,\ldots,1,\ldots,0)$ where the one is in position $j$. While it is cumbersome to carry out this computation, it is the basis for stochastic estimators of the log-determinant. Such estimators have been studied for the trace of a matrix, the trace of the inverse of a matrix and our case, the trace of the logarithm of a matrix. Details on this can be found in \cite{hutchinson1989stochastic} and \cite{bai1996bounds}. The Hutchinson stochastic estimator is described as follows: 1) Let $\bv_j$, $j=1,\ldots,s$ be vectors with entries $P(v^k=1)=1/2$, $P(v^k=-1)=1/2$ independently. 2) Let 
\begin{align}
	\tr \log \bQ \approx \frac{1}{s} \sum_{j=1}^s \bv_j^T \log (\bQ)\bv_j. \label{eq_est_eq}
\end{align}	
As this is a Monte Carlo method, it is possible to compute confidence regions for the estimators, using either Monte Carlo techniques or the Hoeffding inequality (\cite{bai1996bounds}). The Hutchinson estimator was formulated for approximating $\tr \bQ^{-1}$ in which case, we replace the $\log \bQ$ in \eqref{eq_est_eq} with $\bQ^{-1}$.

The Hutchinson estimator requires a lot of random vectors $\bv_j$ to be sufficiently accurate for optimisation. The memory requirements are low, but we may have to wait an eternity for one determinant approximation. The question, then, can we choose the $\bv_j$s in a clever way, so that we require far fewer vectors? 

In recent publications, \cite{bekas2007estimator} and \cite{diagonal_inverse} explored the use of probing vectors for extracting the diagonal of a matrix or its inverse. In the first of these the diagonal of a sparse matrix is extracted, and it is relatively straightforward under mild assumptions to extract the diagonal. The second relies on approximate sparsity of the inverse, where the approximate sparsity pattern of $\bQ^{-1}$ is determined by a power of the original matrix, i.e. $\bQ^p$. Assuming such a sparsity structure, it is possible to compute the probing vectors $\{\bv_j\}_{j=1}^s$ by a neighbour colouring of the graph induced by $\bQ^p$ (see e.g. \cite{iterated_greedy_culb} for a survey on the greedy graph colouring algorithms). A heuristic suggested in \cite{diagonal_inverse} to find the power, $p$ in $\bQ^p$ is to solve $\bQ \bx= \be_j$ and find $p=\min\{d(l,j) | |x_l|<\epsilon\}$ where $d$ gives the graph distance. In our case, we may compute $\log (\bQ) \be_j$ and use the same heuristic. A nice feature is that the probing vectors need not be stored, but may be computed cheaply on the fly. If we pre-compute them, they are sparse, and does not need much storage. Since what we need for each probing vector is $\bv_j^T \log (\bQ) \bv_j$, we observe that the computation is highly parallel with low communication costs. Each node gets one probing vector, and computes $\bv_j^T \log (\bQ) \bv_j$ and sends back the result. In essence, this leads to linear speedup with the amount of processors available with proportionality close to unity.

Next, we need to consider the evaluation of $\log (\bQ) \bv_j$. The matrices we consider have real positive spectrum, and it is possible to evaluate $\log(\bQ)\bv_j$ through Cauchy's integral formula, $$\log(\bQ)\bv_j= \oint_\Gamma \log(x) (z \bI - \bQ)^{-1} \bv_j dz,$$ where $\Gamma$ is a suitable curve enclosing the spectrum of $\bQ$ and avoiding branch cuts of the logarithm. Direct quadrature  over such curves can be terribly inefficient, but through clever conformal mappings, \cite{higham_hale_tref} developed midpoint quadrature rules that converge rapidly for increasing number of quadrature points at the cost of needing estimates for the extremal eigenvalues of $\bQ$. In fact, $\| \log\bQ-f_N(\bQ)\| = \mcO(e^{-2\pi N/(\log(\lambda_{max}/\lambda_{min})+6)})$ with $f_N$ as below. This essentially gives us the rational approximation
\begin{align}
	\log(\bQ)\bv_j \approx f_N(\bQ)\bv_j = \sum_{l=1}^N \alpha_l (\bQ-\sigma_l \bI)^{-1} \bv_j, \,\,\, \alpha_l,\sigma_l \in \mathbb{C}. \label{eq_rat_app}
\end{align}
In effect, we need to solve a family of shifted linear systems to approximate $\log(\bQ)\bv_j$. How we compute $f_N(\bQ) \bv_j$ is problem dependent, but in high dimensions, we usually have to rely on iterative methods, such as Krylov methods. A Krylov subspace, $\mcK_k(\bQ,\bv)$ is defined by $\mcK_k(\bQ,\bv)=\textnormal{span}\{\bv,\bQ \bv, \bQ^2 \bv, \ldots, \bQ^k \bv\}$ and a thorough introduction to the use and theory of Krylov methods can be found in \cite{saad_it_sparse}. Which Krylov method we use is highly dependent on the quality and performance potential preconditioners for the matrix $\bQ$.

While the Krylov method of choice is problem dependent, there are ones that are particularily well suited to compute the approximation in \eqref{eq_rat_app}. These methods are based on the fact that $\mcK_k(\bQ,\bv) = \mcK_k(\bQ - \sigma_l \bI,\bv)$ and after some simple algebra, we obtain the coefficients for the shifted systems without computing new matrix-vector products, see \cite{jeger_shift} and \cite{frommer2003bicgstab} for details. We have employed the method CG-M in \cite{jeger_shift} our computations. One possible difficulty in employing the method is that we have complex shifts - this is remedied by using a variant, Conjugate Orthogonal CG-M (COCG-M), which entails using the conjugate symmetric form $(\overline{\bx},\by)=\bx^T \by$ instead of the usual inner product $(\bx,\by)=\overline{\bx}^T \by$ in the Krylov iterations. See \cite{van1990petrov} for a description of the COCG method.  In practice, little complex arithmetic is needed, since the complex, shifted coefficients are computed from the real ones obtained by the CG method used to solve $\bQ \bx=\by$.

\section{Examples}

In this section we explore how optimisation fares under different conditions. For doing this, we assume essentially the simplest possible model, but we plan use the outlined approach on a seismic case in the future. The model we assume is the SPDE $\tau (\kappa-\triangle)u=\mathcal{W}$, in 2D, which we observe directly. We will explore how optimisation works (on $\tau,\kappa$) for different $\kappa$s and different distance colouring of the graph. Note that the number of colours needed is essentially independent of the granularity of the discretisation: a fine grid yields approximatelly the same number of colours as a coarse grid. The initial suspicion is that for small $\kappa$s, corresponding to long range will be harder to optimise in the following sense: we need more Krylov iterations for the COCG-M routine to converge and we need a larger distance colouring to cover the increasing range, resulting in more probing vectors. We use a modified Newton method for optimisation and compare using the exact determinant to using the approximation outlined above. For the COCG-M method, we use a relative tolerance of $10^{-3}$ for computing $\log (\bQ) \bv_j$. Note that a prior on the parameters will stabilise the results as usual. The results are given in Table \ref{table_kappa}.

\begin{table}[H] 
\begin{centering}
\caption{Optimisation of $(\kappa, \tau)$ for different distance colourings} \label{table_kappa}
\begin{small}
\begin{tabular}{|c|c|c|c|c|c|c|}
\hline  & Exact & 2-dist & 4-dist & 6-dist & 8-dist & 10-dist\\ 
\hline $\kappa=1$ & $( 0.927,1.015)$ & $(1.06,0.98)$ & $(0.933,1.013)$ & $( 0.927,1.015)$ & $\ldots$ & $\ldots$ \\ 
\hline $\kappa=0.5$ & $(0.455, 1.010)$  & $(0.605, 0.961)$  & $( 0.471, 1.005)$ & $( 0.457,1.009)$ & $(0.455,1.010)$ & $\ldots$ \\ 
\hline $\kappa=0.1$ & $(0.0842,1.003)$ & $(0.208,0.940)$ & $(0.122,0.984)$ & $(0.0983,0.996)$ & $(0.0891,1.000)$ & $(0.0859,1.002)$ \\ 
\hline $\kappa=0.05$ & $(0.0398,1.001)$ & $(0.138,0.941)$ & $(0.0762,0.980)$ & $(0.0567,0.992)$ & $(0.0475,0.997)$ & $(0.0434,0.999)$ \\ 
\hline $\kappa=0.01$ & $(0.00644,0.998)$ & $(0.0565,0.947)$ & $(0.0292,0.978)$ & $(0.197,0.987)$ & $(0.0143,0.992)$ & $(0.0117,0.994)$ \\ 
\hline 
\end{tabular} 
\end{small}
\end{centering}
\end{table}

In Table \ref{table_kappa}, $\ldots$ indicates that the optimisation yields the same as the previous entry. We see that as we increase the graph neighbourhood in our colourings, we get results closer and closer to that of using the exact determinant. We also see that the estimates are monotone: $\kappa$ decreases with larger distance colourings and $\tau$ increases. Lastly, we see that some of the estimates are better when using the approximation. This should not necessarily be taken as a good sign, as it may only be a result of approximation errors.

As increasing $k$ in $k$-distance colourings yields better and better approximations, one could be lead to using a "large" $k$ in the entire optimisation, whatever method one uses to determine $k$. Our results indicate, however, that we only need very good approximations in the last iterations of the optimisation procedure. In effect, we may use $2$-distance colourings in the beginning, and go to $5$- or larger -distance colourings in the last steps. Computing the colourings is cheap and one colouring only requires the storage of one vector, so we may store a couple of different colourings and use them as required in the optimisation procedure. 

Lastly, we present an example that cannot be done using black-box Cholesky factorisations. Namely, a 3-D version of the model above with $\kappa=0.5$ and $2$ million discretisation points. We use a $1$-distance colouring for the first iterations and increase to $2$- and $6$-distance colouring in the last iterations. This will give us temporary $k$-distance estimates which we also give. The estimates, progressively from $1$-, $2$ and $6$-distance colouring were $(7.627,0.382), \quad (1.401,0.801),$ and $(0.561,0.988)$. It took $24$ hours to complete the optimisation. From this we conclude - the method is slow, but it can be used for parameter estimation in high-dimensional problems where other alternatives are impossible due to memory limitations. We must, however, be careful so that we have enough colours to capture the essentials of the determinant. The estimated memory use for using Cholesky factorisation in the determinant evaluations is $155$ Gb with METIS sorting of the graph and much higher without. Few computing servers have this amount of memory on a node. The memory consumption for the approximation was $3$ Gb at maximum, and even this may be lowered quite a lot with some clever memory management. In a computing cluster, the time for computing this optimisation would have been much lower due to linear speedup vs. number of nodes.

We have tried the approximation for different types of matrices, and what we found is that the examples above are among the toughest to do determinant approximations on. Since we can do reasonable approximations for this class, we expect that these likelihood evaluations will work  for a large class of precision matrices in use in statistics.

\section{Discussion}

We have showed that the determinant approximations discussed shows promise for likelihood evaluations in models where we cannot perform Cholesky factorisations or Kronecker decompositions of the precision matrices. This may prove useful in high dimensional models where approximate likelihoods are not sufficient for accurate inference. It remains to utilise the approximations on a real world dataset.

\bibliographystyle{apa}

\begin{thebibliography}{}

\bibitem[\protect\astroncite{Bai and Golub}{1997}]{bai1996bounds}
Bai, Z. and Golub, G. (1997).
\newblock {Bounds for the trace of the inverse and the determinant of symmetric
  positive definite matrices}.
\newblock {\em Annals of Numerical Mathematics}, 4:29--38.

\bibitem[\protect\astroncite{Bekas et~al.}{2007}]{bekas2007estimator}
Bekas, C., Kokiopoulou, E., and Saad, Y. (2007).
\newblock {An estimator for the diagonal of a matrix}.
\newblock {\em Applied numerical mathematics}, 57(11-12):1214--1229.

\bibitem[\protect\astroncite{Buland and Omre}{2003}]{bul_omr_linavo}
Buland, A. and Omre, H. (2003).
\newblock Bayesian linearized avo inversion.
\newblock {\em Geophysics}, 68:185--198.

\bibitem[\protect\astroncite{Carlin and Louis}{2000}]{carlin2000bayes}
Carlin, B. and Louis, T. (2000).
\newblock {\em {Bayes and Empirical Bayes methods for data analysis}}.
\newblock CRC Press.

\bibitem[\protect\astroncite{Chen et~al.}{2008}]{chen2008algorithm}
Chen, Y., Davis, T., Hager, W., and Rajamanickam, S. (2008).
\newblock {Algorithm 887: CHOLMOD, supernodal sparse Cholesky factorization and
  update/downdate}.
\newblock {\em ACM Transactions on Mathematical Software (TOMS)}, 35(3):22.

\bibitem[\protect\astroncite{Culberson}{1992}]{iterated_greedy_culb}
Culberson, J. (1992).
\newblock Iterated greedy graph coloring and the difficulty landscape.
\newblock Technical report, University of Alberta.

\bibitem[\protect\astroncite{Davis and Hager}{1999}]{davis1999modifying}
Davis, T. and Hager, W. (1999).
\newblock {Modifying a sparse Cholesky factorization}.
\newblock {\em SIAM Journal on Matrix Analysis and Applications},
  20(3):606--627.

\bibitem[\protect\astroncite{Frommer}{2003}]{frommer2003bicgstab}
Frommer, A. (2003).
\newblock {BiCGStab (l) for families of shifted linear systems}.
\newblock {\em Computing}, 70(2):87--109.

\bibitem[\protect\astroncite{Hale et~al.}{2008}]{higham_hale_tref}
Hale, N., Higham, N.~J., and Trefethen, L.~N. (2008).
\newblock Computing {$A^\alpha$, $\log(A)$} and related matrix functions by
  contour integrals.
\newblock {\em SIAM Journal of Numerical Analysis}, 46:2505--2523.

\bibitem[\protect\astroncite{Higham}{2008}]{Higham:2008:FM}
Higham, N.~J. (2008).
\newblock {\em Functions of Matrices: {Theory} and Computation}.
\newblock Society for Industrial and Applied Mathematics, Philadelphia, PA,
  USA.

\bibitem[\protect\astroncite{Hutchinson}{1989}]{hutchinson1989stochastic}
Hutchinson, M. (1989).
\newblock {A stochastic estimator of the trace of the influence matrix for
  $\{$L$\}$ aplacian smoothing splines}.
\newblock {\em Commun. Stat. Simula.}, 18:1059--1076.

\bibitem[\protect\astroncite{Jegerlehner}{1996}]{jeger_shift}
Jegerlehner, B. (1996).
\newblock Krylov space solvers for shifted linear systems.
\newblock {\em arXiv.org, arXiv:hep-lat/9612014v1}, NA:NA.

\bibitem[\protect\astroncite{Karypis and Kumar}{1999}]{karypis1999fast}
Karypis, G. and Kumar, V. (1999).
\newblock {A fast and high quality multilevel scheme for partitioning irregular
  graphs}.
\newblock {\em SIAM Journal on Scientific Computing}, 20(1):359.

\bibitem[\protect\astroncite{Lindgren et~al.}{2011}]{lindgren2010explicit}
Lindgren, F., Lindstr{\o}m, J., and Rue, H. (2011).
\newblock {An explicit link between Gaussian fields and Gaussian Markov random
  fields: The SPDE approach}.
\newblock {\em Journal of the Royal Statistical Society, Series B}, 5, to
  appear.

\bibitem[\protect\astroncite{McPeters et~al.}{1996}]{mcpeters1996nimbus}
McPeters, R., Aeronautics, U. S.~N., Scientific, S.~A., and Branch, T.~I.
  (1996).
\newblock {\em {Nimbus-7 Total Ozone Mapping Spectrometer (TOMS) data products
  user's guide}}.
\newblock NASA, Scientific and Technical Information Branch.

\bibitem[\protect\astroncite{Rue and Held}{2005}]{rue_gmrf}
Rue, H. and Held, L. (2005).
\newblock {\em Gaussian Markov Random Fields}.
\newblock Chapman {\&} Hall.

\bibitem[\protect\astroncite{Rue and Tjelmeland}{2002}]{rue_tje_fitgmrf}
Rue, H. and Tjelmeland, H. (2002).
\newblock Fitting gaussian markov random fields to gaussian fields.
\newblock {\em Scandinavian Journal of Statistics}, 29:31--49.

\bibitem[\protect\astroncite{Saad}{2003}]{saad_it_sparse}
Saad, Y. (2003).
\newblock {\em Iterative Methods for Sparse Linear Systems, 2nd Ed.}
\newblock SIAM.

\bibitem[\protect\astroncite{Tang and Saad}{2010}]{diagonal_inverse}
Tang, J. and Saad, Y. (2010).
\newblock A probing method for computing the diagonal of the matrix inverse.
\newblock Technical report, Minnesota Supercomputing Institute for Advanced
  Computational Research.

\bibitem[\protect\astroncite{van~der Vorst and Melissen}{1990}]{van1990petrov}
van~der Vorst, H. and Melissen, J. (1990).
\newblock {A Petrov-Galerkin type method for solving Axk= b, where A is
  symmetric complex}.
\newblock {\em Magnetics, IEEE Transactions on}, 26(2):706--708.

\end{thebibliography}

\end{document}